# Using biocompatible materials as substrate coating for electric field enhancement in tip-enhanced Raman spectroscopy

**Fatemeh Sadat Khademi[1], Author**

**Maryam Bahreini[2]*, Advisor**

**Abstract**

In this article, tip-enhanced Raman spectroscopy (TERS) is investigated as a precise method for analysis of biological samples. Using Finite Difference Time Domain (FDTD) simulation, it has been tried to design the required structures for analysis of these samples. At first, by comparing different TERS structures and considering the material, dimensions and other parameters in the simulation, the ideal structure is introduced from a physical point of view and considering the electric field enhancement. In the following, taking into possibility of the effect of environmental and chemical reactions during testing on biological samples, biocompatible materials are used as substrate coating in the simulation, and the effect of using these materials are investigated in comparison with the previous conditions. After performing the simulations, we concluded that Au, Cu and Ag have the highest electric field enhancement, respectively, and the presence of Au next to Cu, also Au and Cu next to Ag, leads to the enhancement of the electric field in them. In the following, we found that the material and thickness of the layer under the coating in the substrate and tip have a great effect on the enhancement. finally, using five biocompatible materials as a coating in the case of using the Au tip and substrate, which creates the most electric field enhancement, we saw that the use of biocompatible materials greatly reduces the enhancement and the use of these five materials does not differ much from each other. Anyway, the use of a 1 nm layer of biocompatible coating creates a much more favorable effect on enhancement than upper thicknesses.

**Introduction**

Tip-enhanced Raman spectroscopy (TERS) is a combination of vibrational Raman spectroscopy and scanning probe microscopy (SPM), which has the chemical sensitivity of Raman spectroscopy and the nanometer spatial resolution of SPM microscopy. After the light hits the material, phenomena including absorption, transmission or scattering may occur. Each of these phenomena can provide basic information about the chemical structure in the material [1]. Raman spectroscopy, which includes the study of Raman scattering, was introduced in 1928 by an Indian scientist named Raman, and today it is a powerful method for chemical analysis and studying the molecular structure of solid, liquid, and gas materials [2, 3]. One of the main limitations of Raman spectroscopy is the weakness of the Raman signal. To overcome this defect, Fleishman et al. introduced surface-enhanced scattering (SERS) in 1974 [4]. In this method, by using the absorption of sample molecules on the rough metal surface, by means of electric field and chemical enhancement, the Raman signal is greatly enhanced [5]. The SERS method largely overcame the problem of the weak Raman signal, but its spatial resolution remained at the same scale as Raman scattering, which was about 200 nm [6]. To overcome this limitation, Wessel proposed in 1985 that by combining the SERS technique and SPM microscopy, nanoparticles in TERS can be located or replaced with a sharp metal tip, and in this method, the spatial resolution can be reduced to 10 nm [7]. In the TERS method, Raman signal enhancement occurs only near a sharp atomic tip, which is usually covered with metals such as Au or Ag and has a size in the range of 10 to 50 nm [8]. In this method, the metal tip is precisely set by the SPM microscope on the surface of the sample that is placed on the metal substrate. When the laser light with the appropriate wavelength is focused on the top of the tip, a strong electromagnetic field is created between the tip and the surface of the substrate, which enhances the Raman signal [6]. In this research, using finite difference time domain (FDTD) simulation by Lumerical software, an attempt has been made to optimize the TERS technique for efficient measurements on biological samples. To perform most of the biomolecular experiments with the TERS method, a strong electrical field enhancement is needed, and to provide this enhancement, metal tips and substrates are usually used. However, metal surfaces are usually not biocompatible

[1] Fatemeh Sadat Khademi, School of physics, Iran University of Science and Technology (e-mail: fs13khademi@gmail.com)
[2] Maryam Bahreini, School of physics, Iran University of Science and Technology (e-mail: m_Bahreini@iust.ac.ir)

[9]. The hydrophobicity of metal surfaces leads to difficulty in deposition and can have a destructive effect on the structure of native protein or nucleic acid. For example, surface-enhanced Raman spectroscopy studies show that the plasmonic metal surface changes the nature of protein and DNA double-strand break[10]. To solve this problem, a thin layer of biocompatible materials is used as a substrate coating [11-13].

In this article first, a comparison is made between the simulation of different structures used in biological experiments by means of TERS method to achieve the appropriate electric field enhancement. In the following, according to the simulations carried out in this field, in order to prevent damage to the sample, biocompatible materials including muscovite mica [10], polyethylene [14], polystyrene [14], polyvinylpyrrolidone(PVP) [15] and poly (N – isopropylacrylamide) (PNIPAM) [15] are used as a thin layer on metal substrates, and their effect on enhancement in different modes will be compared.

**Simulation method**

In this article, finite difference time domain (FDTD) simulation by Lumerical software is used. In 1966, Yi proposed a new method for solving Maxwell's equations based on the finite difference method. FDTD is one of the most popular and widely used numerical methods for the explicit solution of Maxwell's equations, which has become very popular for various reasons, including the ability to solve equations with a large number of unknowns, being strong and generality. FDTD is a completely vector method that provides the information of the frequency and time domains to the user and provides a different view of various problems and applications in the science of electromagnetics and photonics[16].

The desired geometry for simulation in this research is shown in the fig.1. In this simulation, the tip is in the form of a cone with length L, whose apex is a circle with radius r and its end is a circle with radius D/2, and it is located at a distance d from the surface of the substrate with thickness t. Also, a linearly polarized light that makes an angle θ with the axis of the beam is used to create Raman scattering [17-19].

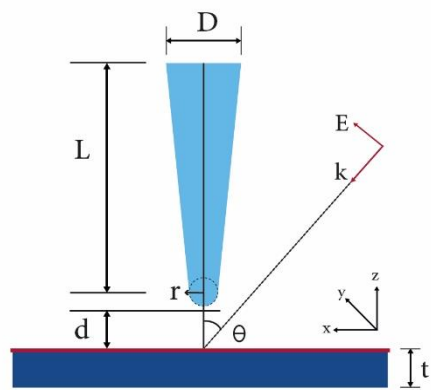

**Fig.1** The parameters in simulation of the geometry of tip and the substrate

**Results and discussion**

By reviewing the articles related to TERS experiments in biological fields, it can be seen that Au, Ag, and Cu are usually used as the tip and substrate to achieve high electrical enhancement[20, 21]. each of these metals has advantages and disadvantages. In short, all three metals offer a favorable electric field enhancement, however, the high cost of Au, rapid oxidation of Ag exposed to air[22], and the high reactivity of Cu and as a result of its low chemical stability have caused the use of these three materials to face challenges [15, 19, 23-25]. In this section, the simulation of different states of use of these three metals is examined. In the following, due to the fact that TERS experiments usually use Si tips coated with different metals[26-28], the effect of using Si in the bottom layer of the tip will be investigated in the simulation [29]. Also, in order to prevent damage to the sample, in the final simulations, biocompatible materials are used as a thin layer on metal substrates, and its effect on strengthening in different states is compared.

It should be noted that in all TERS simulations, a tip with a length of 300 nm, whose tip is a circle with a radius of 25 nm, and which is located at a distance of 0.5 nm from the surface of the substrate with a thickness of 55 nm, has been used. The amplitude of the electric field of the incident light is also considered to be 1 (v/m).

1. *Using the tip and substrate with the same material, including three modes of using Au, Ag and Cu*

In this part, the electric field enhancement is investigated in three cases of using the tip and substrate with the same material of Au, Ag and Cu. The simulation of the Au tip and substrate can be seen in Fig.2. According to the diagram in Fig.3, the maximum enhancement of the electric field is at the wavelength of 633 nm for the tip and substrate of Au, Cu and Ag, respectively. At the wavelength of 785 nm, all three structures show almost the same reaction, and at the wavelength of 532 nm, more enhancement can be obtained by using Ag. Au and Ag, due to their high free electron density, and Cu and Au, due to their low extinction coefficient (the imaginary part of the dielectric constant), are the best options for creating an electric field enhancement suitable for conducting biological experiments by the TERS method. Having a high extinction coefficient, Ag absorbs a large share of incoming light compared to Au and Cu, thereby reducing the electric field enhancement [23, 30, 31].

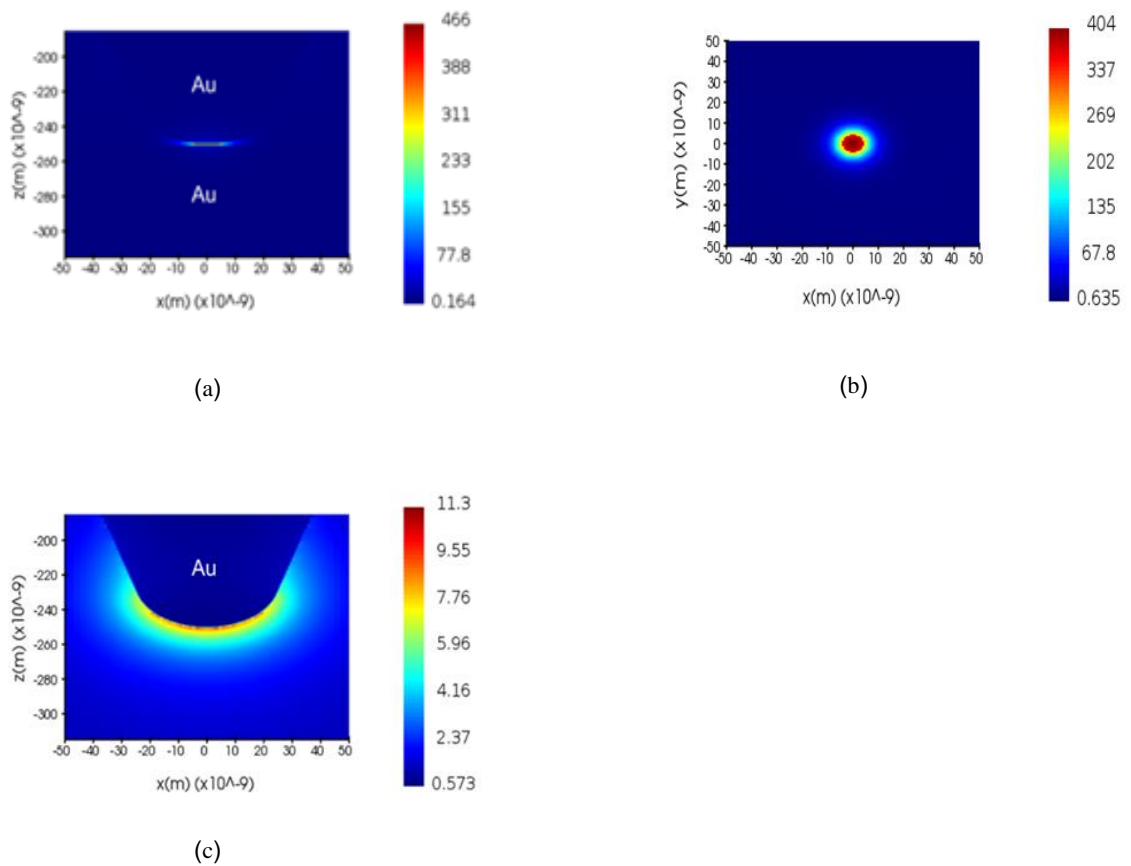

**Fig.2** Distribution of the electric field in the case of: (a) using the tip and Au substrate along the y-axis (b) state a along the z-axis (c) state a without considering the substrate

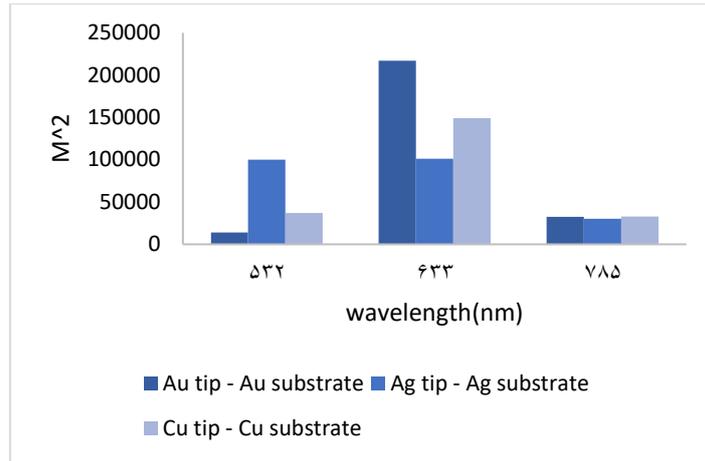

Fig.3 Comparison diagram of electric field enhancement in terms of wavelength for the use of Au, Ag and Cu tip and substrate

2. *using of different metals for the tip and substrate, including Au, Ag and Cu*

By comparing the different modes of using different metals for the tip and substrate, it can be seen in the table1 and diagram of Fig.4, at the wavelength of 633 nm, the maximum enhancement of the electric field is obtained. in each case, at the wavelength of 785 nm, the material of tip and the substrate do not have a special effect on enhancement, and at the wavelength of 532 nm, when Ag is used, especially as a tip, better enhancement is created. In the case of using Au tip and substrate, the maximum electrical enhancement can be obtained, and the combination of Au with Ag and Cu will not increase the enhancement. When using a Cu tip or substrate, the use of Au next to it increases and the use of Ag decreases the enhancement of the electric field. and finally, when using Ag tip or substrate Using Au or Cu next to Ag, increases its electric field.

Table 1 Comparison of the electric field enhancement ($|M|^2$) and the y component of the electric field ($|M|$) in terms of wavelength for the modes of using the tip and substrate with different types of Au, Cu and Ag

| type | λ (nm) | $|M|$ | $|M|^2$ |
|---|---|---|---|
| Au tip-Ag substrate | 532 | 156 | 24336 |
| | 633 | 393 | 154449 |
| | 785 | 175 | 30625 |
| Au tip-Cu substrate | 532 | 139 | 19321 |
| | 633 | 430 | 184900 |
| | 785 | 181 | 32761 |
| Ag tip-Au substrate | 532 | 250 | 62500 |
| | 633 | 398 | 158404 |
| | 785 | 178 | 31684 |
| Ag tip-Cu substrate | 532 | 280 | 78400 |
| | 633 | 360 | 129600 |
| | 785 | 179 | 32041 |
| Cu tip-Au substrate | 532 | 169 | 28561 |
| | 633 | 426 | 181476 |
| | 785 | 179 | 32041 |
| Cu tip-Ag substrate | 532 | 204 | 41616 |
| | 633 | 349 | 121801 |
| | 785 | 175 | 30625 |

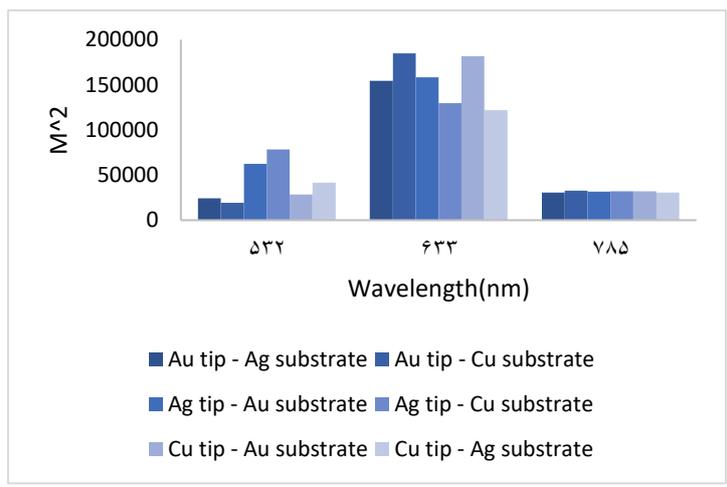

Fig.4 Comparison diagram of electric field enhancement in terms of wavelength for the use of tip and substrate with different types of Au, Cu and Ag

3. *using Au and Cu metal coating on the tip and substrate with Cu and Ag materials*

According to the diagram in Fig.5, the use of Au coating for the Ag substrate has a greater effect on the increase of enhancement than the other two cases of using the coating on the substrate. However, the Au coating next to the Cu substrate creates stronger electric field enhancement.

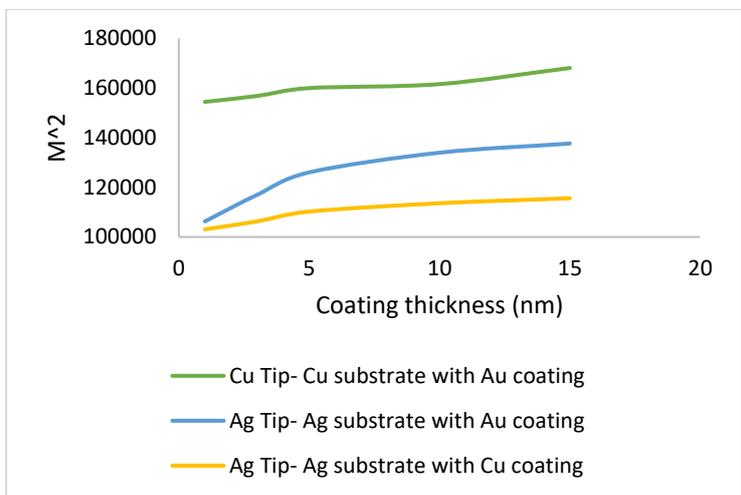

**Fig.5** The comparison diagram of the electric field enhancement in terms of wavelength for the case of using the substrate coating for Cu and Ag substrates.

According to the diagram in Fig.6, when using the coating for the tip, the use of Au cover for the Ag tip has a greater effect on increasing the enhancement than the other two cases. However, the Au coating next to the Cu tip creates more electric field enhancement.

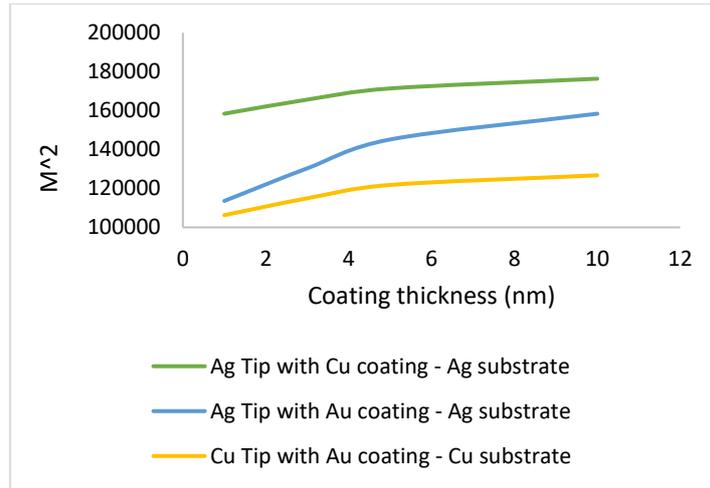

Fig.6 Comparison diagram of electric field enhancement in terms of wavelength for the case of using tip coating for tip and Cu and Ag substrates.

4. *Investigating the underlying layer in the substrate using glass ($SiO_2$) instead of the Cu substrate in simulating the Cu substrate with 3 nm Au coating and Cu tip*

According to the simulation of Fig.7 and using glass instead of Cu in the bottom layer, it can be seen that the material of the bottom layer of the substrate has a great effect on enhancement the electric field.

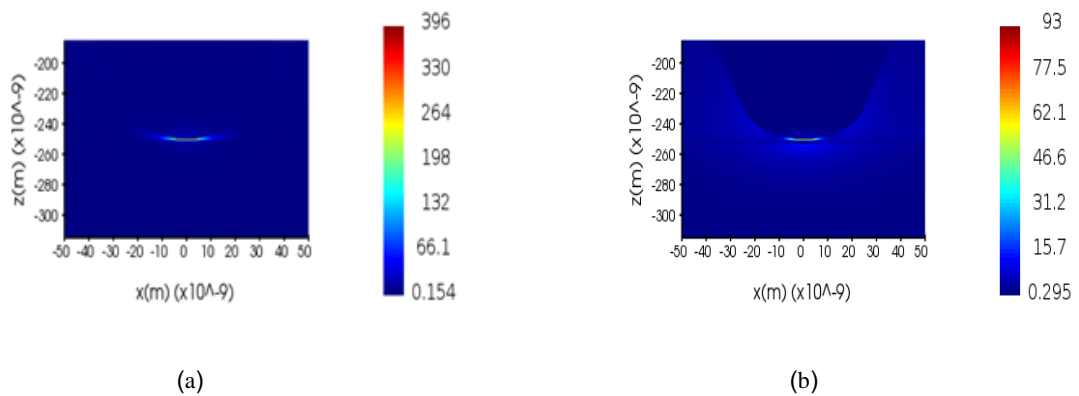

(a)                                         (b)

**Fig.7** The distribution of the y component of the electric field in two cases: (a) using a Cu substrate with 3 nm Au coating and Cu tip and (b) using glass ($SiO_2$) instead of the Cu substrate in case a.

5. *Using different thicknesses of Cu substrate in the case of Cu substrate with 3 nm Au coating and Cu tip*

In this part, different thicknesses of Cu have been used as the bottom layer of the substrate in order to check whether the increase in the electric field enhancement created when using the Au coating on the substrate is only due to the use of the coating or due to the interaction between the coating and the substrate is related. According to the diagram in Fig.8, in the thicknesses of 15 to 35 nm of Cu, the enhancement increases with a large slope, and in the thicknesses of 35 to 55 nm, the enhancement of the electric field decreases with a very small slope.

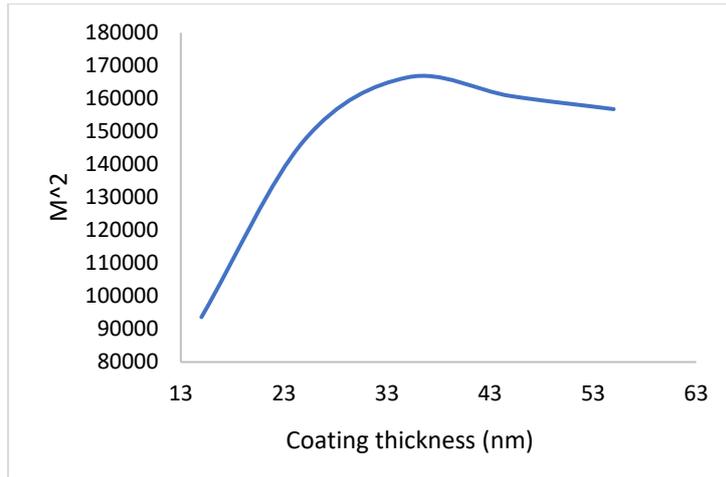

Fig.8 The diagram of the electric field enhancement according to the thickness of the Cu substrate in the case of Cu substrate with 3 nm Au coating and Cu tip.

6. *The effect of increasing the distance between the tip and the substrate on the electric field enhancement*

In this simulation, Au tip and substrate are used at different distances from each other and a laser with a wavelength of 633 nm is used. Due to the diagram in Fig.9, it can be seen that when the distance between the tip and the substrate increases between 0.5 and 3 nm, the electric field enhancement decreases with a large slope and after the distance of 3 nm, until If the substrate is placed at an infinite distance from the tip, or in other words, when the tip is used without the presence of the substrate, the electric field enhancement decreases with a very small slope. Also, it can be seen in the enhancement diagrams according to the location of the tip, at the distance of 0.5 and 1 nm, the highest enhancement is at the wavelength of 633 nm. at the distances greater than 2 nm, the lowest amount of electric field enhancement is created at the apex of the tip. It can be Therefore, it is suggested to create a higher enhancement at the tip apex, the tip and the substrate should be placed at a distance of 0.5 or at most 1 nm from each other.

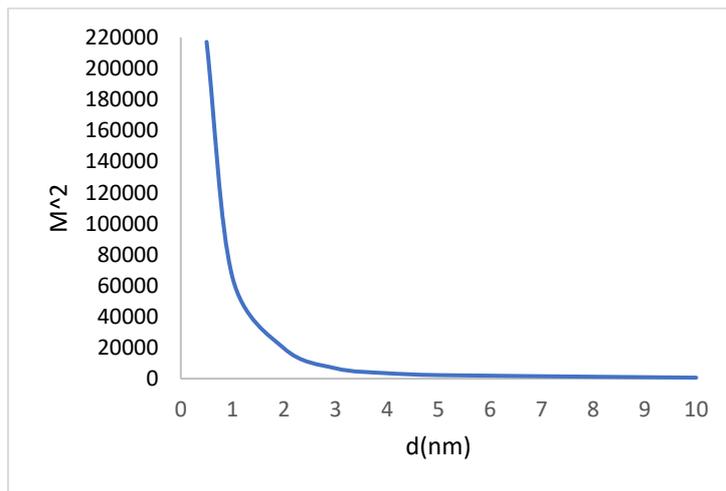

Fig.9 Electric field enhancement diagram according to the distance between the tip and the substrate

7. *Useing of Si tip with Au coating*

In the experimental tests with TERS method, Si tips which are coated with the metal are usually used. In this section, the difference between the use of Si tip with different thicknesses of Au coating compared to the use of pure Au tip is examined. In all these cases, the tip radius is considered to be 25 nm. As can be seen in the diagram in Fig.10, from 1 to 5 nm thickness, the enhancement decreases with increasing thickness. that is, 1 nanometer thickness provides better enhancement than 5 nanometer state. From the thickness of 5 to 25 nm, the enhancement increases and reaches its maximum value in the case of using pure Au tip. It should be noted that in all these cases, a Au substrate with a thickness of 55 nm and a distance of 0.5 nm from the tip was used. Fig.11 also shows the electric field distribution along the z axis in these two cases.

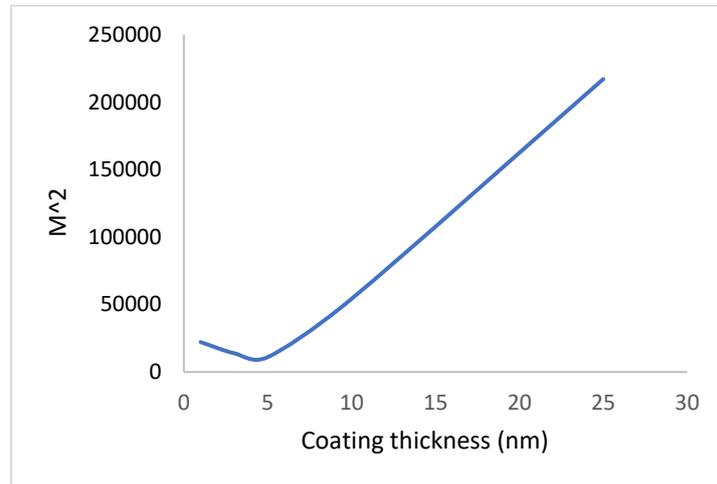

Fig.10  Electric field enhancement diagram according to the thickness of the Au coating on the Si tip

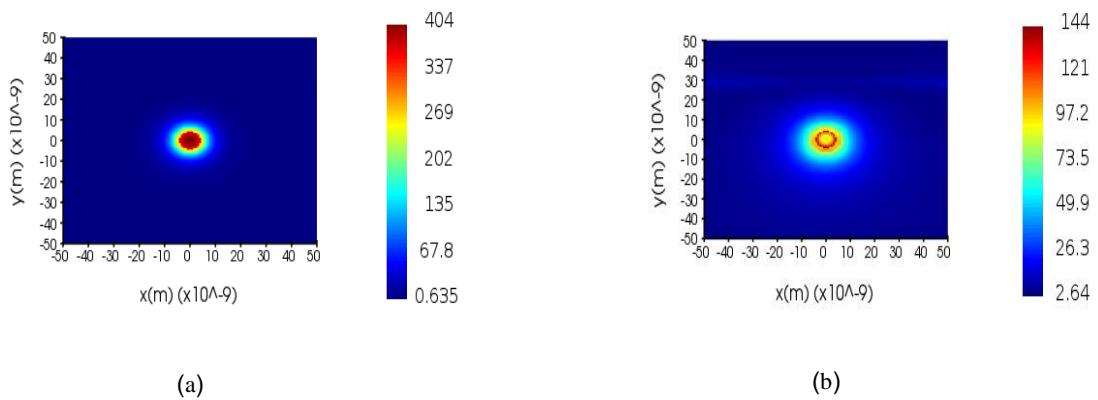

(a)            (b)

**Fig.11** Electric field distribution along the z-axis in the case of: (a) using Au tip and substrate (b) using Si tip with 1 nanometer Au coating and Au substrate

8. *Useing of biocompatible materials as substrate coating*

To perform most of the biomolecular experiments with the TERS method, a strong field enhancement is needed, and to provide this enhancement, metal tips and substrates are usually used. However, metal surfaces are usually not biocompatible. To solve this problem, a thin layer of biocompatible materials is used as a substrate coating [10].

In this section, as can be seen in Table 4, from five biocompatible materials, muscovite mica, polyethylene (PE), PNIPAM, polystyrene and polyvinyl pyrrolidone (PVP) that using as substrate coating with different thicknesses in the case of using the tip and the Au substrate, whose electric field in the y-axis is 466 (visible in the previous parts). The materials used in this simulation are defined in the software, according to the refractive index (real part (n) and imaginary part (k)) at the wavelength of 633 nm. According to the results, muscovite mica, polystyrene, PVP, polyethylene and PNIPAM provide higher enhancement. however, the amount of enhancement in these five materials is slightly different from each other. Also, according to the graphs in Fig.12, when using 1 to 3 nm of the coating for the substrate, the electric field enhancement decreases strongly and with a large slope, and in thicknesses greater than 4 nm, the enhancement becomes almost constant. . The electric field component is reduced from 466 to 282 in the best case, that is, it is almost halved and faces a significant reduction, but according to the cases reviewed in Table2, the use of a 1 nm layer Biocompatible coating creates a much more favorable effect on enhancement than other modes.

**Table 2** Electric field enhancement according to the thickness of the coating of biocompatible materials on the Au substrate

| Substrate coating | index of refraction | thickness (nm) | 1 | 3 | 5 | 10 |
|---|---|---|---|---|---|---|
| muscovite mica | n = 1/6 | $|M|$ | 282 | 125 | 97.5 | 65.1 |
| | k = 0 | $|M|^2$ | 79524 | 15625 | 9506.25 | 4238.01 |
| PNIPAM | n = 1.5012 | $|M|$ | 266 | 137 | 87.5 | 57.4 |
| | k = 0.0022 | $|M|^2$ | 70756 | 18769 | 7656.25 | 3294.76 |
| polyethylene | n = 1.519 | $|M|$ | 269 | 139 | 89.3 | 58.7 |
| | k = 0.002 | $|M|^2$ | 72361 | 19321 | 7974.49 | 3445.69 |
| Polystyrene | n = 1.5875 | $|M|$ | 280 | 149 | 96.3 | 64.2 |
| | k = 0 | $|M|^2$ | 78400 | 22201 | 9273.69 | 4121.64 |
| PVP | n = 1.5252 | $|M|$ | 270 | 140 | 89.9 | 59.2 |
| | k = 0.0018 | $|M|^2$ | 72900 | 19600 | 8082.01 | 3504.64 |

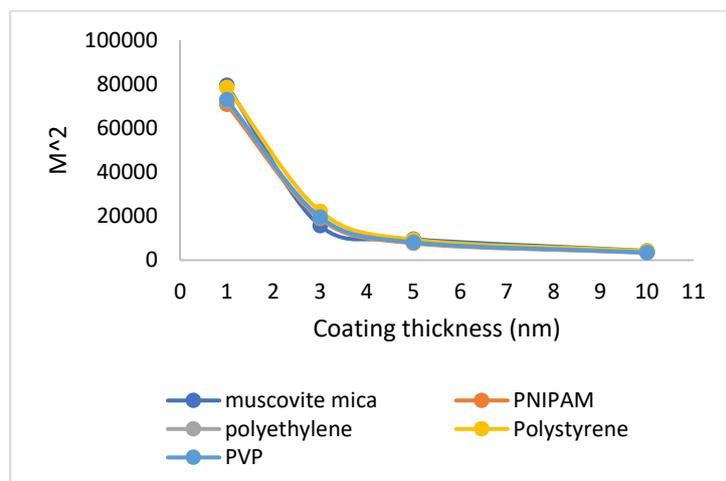

**Fig.12** The diagram of electric field enhancement according to the coating thickness of biocompatible materials on the Au substrate

**conclusion**

In this article, it was observed that at the wavelength of 633 nm, respectively, Au, Cu and Ag have the highest electric field enhancement. When using a Cu tip or substrate, it was observed that the use of Au next to it, increases and the use of Ag decreases the enhancement of the electric field, and when using Ag tip or substrate, the use of Au or Cu, along with it, increases the enhancement of the electric field. Considering that from an economic point of view, it is more economical to use a thin layer of Au as a coating on a tip or a substrate made of Cu or Ag, instead of using a tip and a substrate that is completely made of Au. in addition, Au and Cu coating for Ag reduces corrosion against air, and Au coating on Cu reduces the chemical reaction when using it. in the following, the effect of Au coating on the Cu tip and substrate and the effect of Au coating and Cu on the Ag tip and substrate were investigated at the wavelength of 633 nm and it was observed that the use of Au coating for Ag has a greater effect on increasing the enhancement than the two cases of Au coating for Cu and Cu coating for Ag. However, the Au coating next to the Cu substrate creates a greater electric field enhancement. Further, it was observed that the use of the same coating for the tip and the substrate is not very important. But according to the simulations, due to the accumulation of more charge in the sharp places and as a result, the density of free electrons in these places, the same thickness of the coating on the tip provides higher enhancement than the coating on the substrate. To investigate the effect of the material of the bottom layer of the substrate when using the coating on it, glass was used instead of Cu and it was observed that the material of the bottom layer of the substrate has a great effect on strengthening. In the following, it was investigated that different thicknesses of the lower layer of the substrate also have a very different effect on the enhancement.

In the next simulation, the effect of increasing the distance between the tip and the substrate on the enhancement of the electric field was investigated and it was suggested that the tip and the substrate should be placed at a distance of 0.5 or at most 1 nm from each other to create a higher enhancement at the top of the tip.

According to the fact that in experimental tests with TERS method, Si tips which are coated with the other metal are usually used, In the next simulation the difference of using a Si tip with different thicknesses of Au coating was investigated in comparison with the case of using a completely Au tip.

In the end, due to the fact that metal surfaces are usually not biocompatible, five biocompatible materials were used as substrate coating with different thicknesses in the case of using tip and Au substrate, and it was observed that muscovite mica, polystyrene, PVP, polyethylene, and PNIPAM create higher enhancement respectively. however, the amount of enhancement in these five materials is slightly different from each other. The electric field component in the best case is reduced from 466 to 282. that is, it is almost halved and faces a significant reduction. However, the use of a 1 nm layer of biocompatible coating has a much more favorable effect on enhancement than the other states.